\documentclass[12pt]{article}
\usepackage{a4}
\usepackage{graphicx}
\usepackage{amsfonts}
\usepackage{amscd}
\usepackage{latexsym}
\usepackage{epsf}
\usepackage{amsmath}
\numberwithin{equation}{section}
\newcommand{\uv}{ultraviolet }
\newcommand{\ew}{electro weak }

\newcommand{\nn}{\nonumber}
\newcommand{\MVAR}[2]{#1_#2}
\newcommand{\Mvariable}[1]{\MVAR #1}
\newcommand{\fr}{\frac}
\newcommand{\pd}{\partial}

\newcommand{\vp}{\varphi }

\newcommand{\half}{\frac{1}{2}}

\begin{document}
\title{ Vacuum polarization by a magnetic flux  of special rectangular form}
\author{ {\sc I. Drozdov}\thanks{e-mail: 
Igor.Drosdow@itp.uni-leipzig.de} \\ 
\small University of Leipzig, Institute for Theoretical Physics\\ 
\small Augustusplatz 10/11, 04109 Leipzig, Germany}
\maketitle
\begin{abstract}
We consider the ground state energy of a spinor field in the
background of a square well shaped magnetic flux tube. We use the zeta-
function regularization and express the ground state energy as an
integral involving the Jost function of a two dimensional scattering problem.
We perform the renormalization by subtracting the contributions from first
several heat kernel coefficients. The ground state energy is presented as a
convergent expression suited for numerical evaluation. 
We discuss corresponding numerical calculations.
Using the uniform asymptotic expansion of the special functions entering the
Jost function we are able to calculate higher order heat kernel coefficients. 
\end{abstract}
 
\section{ Introduction }
Since the classical work \cite{casimir} of H.B.G. Casimir, where the
energy of vacuum polarized by two conducting planes was calculated, a
number of similar problems was investigated for various external
conditions (boundary conditions, background potentials etc.), see, for
example, the recent review \cite{Bordag:2001qi} or the books
\cite{most,milton} .  No general rule for the dependence of the vacuum
energy on the background properties has been found so far. In
particular, it is unknown how to forecast the sign of the energy.

An interesting problem  is the calculation of the vacuum energy of a
spinor field in the background of a magnetic field. This problem was
first considered as early as in 1936 by W. Heisenberg and H. Euler
\cite{euler} who were interested in the effective action in the
background of a homogeneous magnetic field.

Presently, the interest is shifted to string like configurations.  In
\cite{Groves:1999ks} the contribution of the fermionic ground state
energy to the stability of \ew strings was addressed. In
\cite{Diakonov:2002bx} the gluonic ground state energy in the
background of a center-of-group vortex in QCD had been considered.

A very special inhomogeneous reflectionless magnetic background is for
example the model for domain wall \cite{dunne}. In this case, the
equation for the vacuum fluctuations was easy to solve
analytically. However, the final expression for the effective action
involves an additional integration over a momentum, thus bringing no
real simplifications.

In QED, for the background of a flux tube with constant magnetic field
inside the ground state energy was calculated in \cite{borkir2}. Here
it turned out to be negative remaining of course much smaller than the
classical energy of the background field.

An interesting approach is that used in \cite{fry} where the issue of
the sign and of bounds on fermionic determinants in a magnetic
background had been considered.

The remarkably simple case of a magnetic background field concentrated
on a cylindrical shell (i.e., a delta function shaped profile) was
considered in \cite{scand2}. Here different signs of the vacuum energy
turned out to be possible in dependence on the parameters (radius and flux).  

In general, the issue of stability of the strings is of interest. In
\ew theory, and in QED in particular, the coupling is small. Hence the
vacuum energy as being a one loop correction to the classical energy
is suppressed by this coupling. While in QED a magnetic string is
intrinsically unstable, in \ew theory there are unstable and stable
configuration known \cite{Achucarro:1999it}. Here quantum corrections
may become important for the stability. A question of special interest
is whether a strong or singular background may have a quantum vacuum
energy comparable to the classical one.

For the calculation of ground state energy, it is necessary to
subtract the \uv divergences. This procedure is in general well known
and we follow \cite{}. Roughly speaking one has to subtract the
contribution of the first few heat kernel coefficients ($a_0$ through
$a_2$ in the given case of (3+1) dimensions). After that one can
remove the intermediate regularization and one is left with finite
expressions. However, in order to obtain these finite expressions in a
form suitable for numerical calculations one has to go one more step.
As described in \cite{}, see also below in section 2, one has to add
and to subtract of a certain part of the asymptotic expansion of the
integrand.

In the present paper we generalize the analysis done in \cite{borkir1}
to a rectangular shaped magnetic background field. So we consider the
vacuum energy of a spinor field in QED for a background as given by
Eqs. (\ref{fieldb}, \ref{h_r1}). This problem is technically more involved
and allows progress in two directions. First, it allows to refine the
mathematical and numerical tools for such problems and, second, it
allows to address the question how the vacuum energy behaves for an
increasingly singular background (making the rectangle narrower).  So
this model interpolates to some extend between the flux tube with
homogeneous field inside in \cite{borkir1} and the delta shaped one in
\cite{scand2}.

The basic principles of this procedure are described briefly in
Sec. 2.  Namely it starts with the well known zeta functional
regularization.  The regularized ground state energy is represented as
a zeta function of a hamiltonian spectrum and treated in termini of
heat kernel expansion.  The representation of regularized ground state
energy as an integral of the logarithmic derivative of the Jost
function for spinor wave scattering problem on the external magnetic
background is obtained.  The Jost function is obtained from the exact
solutions of Dirac equation, derived in Sec. 3.  The explicit form of
the exact and asymptotic Jost function is considered in Sec. 4.  In
Sec.5 the representation useful for further numerical evaluations is
derived.  Sec. 6 is devoted to the calculation of the heat kernel
coefficient $a_{5/2}$ and the Sec. 7 contains some numerical
evaluations of the ground state energy.  The divergent part is
identified as that part of the corresponding heat kernel expansion
which does not vanish for large $m$ (mass of the quantum spinor
field). After the subtraction of this divergent part the remaining
analytical expression must be transformed in order to lift the
regularization.  A part of the uniform asymptotic expansion of the
Jost function is used for this procedure.  Finally the analytical
expression for the ground state energy has been evaluated numerically.

Throughout the paper $\hbar=c=1$ is used. 

\section{The renormalized ground state energy}
Consider a spinor field on a magnetic background $\vec{B}$ of the form\\
\begin{equation}
 \vec{B}= \fr{\Phi}{2 \pi} {h(r)}\vec{e_z}
 \label{fieldb}
 \end{equation}
where $\Phi$ is constant defining the magnetic flux, $\vec{e_z}$ is a
unit vector in the cylindrical coordinate system ($ r, \phi, z$).  The
Lagrangian is
\begin{equation}
 {\cal L} = - \fr{1}{4}F_{\mu\nu}F^{\mu\nu}+\bar{\psi} \left[ i\gamma^\mu (
 \pd_\mu+e A_{\mu})-m \right]\psi.
\end{equation}
The potential $\vec{A}(r)$
\begin{equation} 
\vec{A}= \fr{\Phi}{2 \pi} \fr{a(r)}{r}\vec{e_\vp}
\label{potential_A}
\end{equation}
possesses cylindrical symmetry and the radial part
 $a(r)$ is taken to be\\
\hspace*{1cm}
\begin{equation}
a(r)= \left\{\begin{array}{c} 0, \hfill r< R_1, R_2\\
 \fr{1}{\kappa}(\fr{r^2}{R_1^2}-1)\ R_1\le r \le
 R_2 \\ 1, \hspace{\fill} r > R_1, R_2\end{array}\right. \\
\label{a_r1}
\end{equation}
The profile function for the magnetic field is\\
 \hspace*{2cm}
 \begin{equation}
\ \ h(r)=\fr{1}{r}\fr{\pd}{\pd r}a(r)=\left\{\begin{array}{c} 0, \hfill r< R_1, R_2\\\fr{2}{\kappa R^2_1} \hspace{\fill} R_1\le r \le
 R_2 \\ 0, \hspace{\fill} r > R_1, R_2\end{array}\right.  
\label{h_r1}
\end{equation}
with $\kappa= \fr{R_2^2-R_1^2}{R_1^2}$.The shape of the
background can be interpreted geometrically as an infinitely long flux
tube empty inside.

We use the zeta-function regularization for the vacuum energy $E_0$.
Since the background is static, the following representation holds,
\begin{equation}
 E_0(s) =
 -\half\mu^{2s}\sum\limits_{(n)}\omega_{(n)}^{1-2s},\end{equation}
where
\[  \omega_{(n)}= \epsilon_{(n)}=\sqrt{\vec{p}^2+m^2} \]
are eigenfrequences resp.  eigenvalues of energy for one particle
states, (the spectrum of the corresponding hamiltonian, see the next
section), $\mu $ is a parameter with the dimension of mass which is
introduced to keep the correct dimension of energy in this expression.

For technical reasons we assume that the system is contained in a
large but finite cylinder of radius $R$ in order to have discrete
eigenvalues in the transversal directions. Because of the
translational invariance along the z-axis we can separate the
$z$-component of momentum,
 \begin{equation}
 E_0(s)=-\frac{1}{2}\mu^{2s}\int\limits_{-\infty}^{\infty} \frac{dp_z}{2\pi}
 \sum\limits_{(n)} (p_z^2+k_{(n)}^2+m^2)^{\half-s}.
 \end{equation}
Here is it necessary to remark that $E_0$ has the meaning of density per
unit length. Integrating out $p_z$ we get  
\begin{equation}
E_0(s)=-\mu^{2s}\frac{1}{4\sqrt{\pi}} 4\frac{\Gamma(s-1)}{\Gamma(s-\half)}
\sum\limits_{(n)} (k_(n)^2+m^2)^{1-s} .
\label{e0sumk}
\end{equation}
In this expression a factor 4 resulting from the summation over spin
states and over the sign of the one particle energies appeared. The
remaining sum over $(n)$ is over the eigenvalues $k_(n)$ of a two
dimensional wave equation for one mode in the perpendicular plane.
The further transformation of this expression is described more
detailed in \cite{borkir1}, \cite{kirsten}, \cite{BEK}. The first step
is based on the property of logarithmic derivative of eigenfunctions
with respect to momentum k. The contour integral,
 \[ 
\int\limits_\gamma\frac{dk}{2\pi i}(k^2+m^2)^{1-s}
 \partial_k \ln\phi_l(kr), \]
is equal to the sum over the spectrum, $\sum\limits_{(n)}
(k_n^2+m^2)^{1-s}$, if the contour $\gamma$ encircles the real
positive axis.  Then (\ref{e0sumk}) becomes
 \begin{equation}
E_0(s)=-\mu^{2s}\frac{1}{\sqrt{\pi}}
\frac{\Gamma(s-1)}{\Gamma(s-\half)} \sum\limits_{l=-\infty}^{\infty} 
\int\limits_\gamma\frac{dk}{2\pi i}(k^2+m^2)^{1-s} \partial_k \ln\phi_l(kr).\\
\end{equation}
Here $\phi_l(r)$ are the eigenfunctions of hamiltonian
$\hat{H}$ labeled by the orbital momentum $l$.

In the present formalism instead of wave functions $\phi_l(r)$ we use
the corresponding Jost functions $f_l(k)$ which contain all necessary
informations on the spectrum of scattering problem.  For the space
domain outside of cylindrical magnetic background, at $ r > R_2 $, the
wave functions can be chosen as a linear composition of Hankel
functions
 \begin{equation}
 \phi_l(kr) = \half[ \bar{f_l}(k)H^{(1)}_l(kr)+f_l(k) H^{(2)}_l(kr)],\\
 \label{jost_lk}
 \end{equation}
where $\bar{f}_l(k)$ and $f_l(k)$ are the corresponding Jost
functions.
 
Now after the subtraction of contribution from empty Minkowski space
(identified as the term divergent at $R\rightarrow\infty$, which is
independent of the background and corresponds to the heat kernel
coefficient $a_0$,(\ref{a_0}) below), and rotation of the path towards
the imaginary positive axis one obtains \cite{borkir1}
\begin{equation}
E_0(s)= \\ -\frac{1}{\sqrt{\pi}}
\frac{\Gamma(s-1)}{\Gamma(s-\half)}\frac{\sin\pi s}{\pi} 
\sum\limits_{l=-\infty}^{\infty} \int\limits_m^{\infty} dk
(k^2-m^2)^{1-s} \partial_k \ln f_l(ik),
\label{e0_of_lnf}
\end{equation}
which is a very useful representation of the regularized ground state
energy. A merit of this representation is the absence of oscillations
of the integrand for large arguments\footnote{Note that bound states
are also taken into account in $E_0$ (\ref{e0_of_lnf}). In the
considered problem bound states appear if the flux is larger than one
flux unit. Strictly speaking, these are zero modes located on the
lower end of the continuous spectrum, i.e. at k=0 (this is known from
\cite{aharonov_casher}).}.

To discuss divergences in $E_0$ we need a more general setting.
$E_0(s)$ can be expressed through the zeta-function of the associated
differential operator $ \bf{\hat{P}}$
 \begin{equation} 
 E_0(s) =-\frac{\mu^{2s}}{2}\zeta_{(\bf{\hat{P}})}(s-\half).
 \label{e0_zeta}
 \end{equation}
$\zeta_{(\bf{\hat{P}})}(s) $ admits an integral representation 
 \begin{equation}
  \zeta_{\bf{\hat{P}}}(s)=\int\limits_{0}^{\infty} dt
\frac{ t^{s-1}}{\Gamma(s)}K(t)
\end{equation} 
where the kernel of this integral representation $K(t)= \sum\limits_n
e^{-t(k_{(n)}^2+m^2)}$ is the so called heat kernel.  The associated
``heat kernel expansion'' for the kernel $K(t)$ of this integral
representation at $t\rightarrow 0$ is \cite{gilkey}
\begin{equation}
K(t)\sim \fr{e^{-tm^2}}{(4\pi t)^{3/2}} \sum\limits_{n\ge 0} a_n t^n \ .
\label{heat_kt_sum}
\end{equation}
In accordance to the heat kernel expansion, the coefficients $a_n$
occurred in (\ref{heat_kt_sum}) must be for our background \cite{gilkey}:
\begin{equation}
a_0 = V
\label{a_0}
\end{equation}
is simple an (infinite) volume of configuration space; the part proportional
to it has been even dropped before during the renormalization procedure as a
contribution of the empty Minkowski space. All other coefficients up to $a_2$
for this background must be zero through dimensional reasons and requirement
of gauge invariance, (for details see the paper \cite{gkv}). 
\begin{equation}
a_{1/2}=a_1=a_{3/2}=0.
\end{equation}
The first nonzero coefficient is known to read 
\begin{equation}
a_2=\fr{2}{3}F^2_{\mu\nu}
\label{coeff_a_2}
\end{equation} 
and only the term with $n=2$ contributes to the ``divergent'' part of
energy \cite{scand2}, \cite{borkir1}. It will be shown below that only the
contribution proportional to $a_2$ in $E_0(s)$ contains a simple pole at $s
\rightarrow 0 $ and in the case of pure magnetic background is
proportional to the classical energy 
\begin{equation}
E^{class}=\half\vec{B}^2 =\fr{8 e^2}{3}a_2 \ 
\end{equation}

In the paper \cite{borkir1} it has been found, that the regularized
ground state energy of the magnetic flux tube of finite radius $R$
behaves as $~\fr{R^{-3}}{m}$ at $R \rightarrow \infty$ which
corresponds to the contribution proportional to $a_{5/2}$.  It holds
in accordance to the observed fact that in the case of non-smooth
backgrounds the heat kernel coefficients with half-integer number
starting from some value corresponding to the smoothness-class of the
background are different from zero.  It occurs namely in our case,
where the background potential $\vec{A}(r)$ is continous and the
magnetic field $h(r)$ has a discontinuity. The coefficient $a_{5/2}$
is nonzero in our case, see below Sec.6. 
The next terms of this asymptotic may be $~\fr{R^{-4}}{m^2}$
corresponding to $a_3$, and $a_4$ delivering terms
$~\fr{R^{-6}}{m^4}$.  

We define the renormalized ground state energy as
\begin{equation}
 E^{ren}=E_0-E^{div },
 \label{e0-ediv}
\end{equation}
where $E^{div }$ is obtained from the heat kernel expansion
(\ref{heat_kt_sum}) and $E^{ren}$ fulfills the normalization condition
\cite{BKEL}
 \begin{equation}
 \lim E^{ren}=0\ \ at\ \ m \rightarrow \infty .
 \label{norm_cond}
 \end{equation}
Inserting the heat kernel expansion (\ref{heat_kt_sum}) into the zeta
function and using (\ref{e0_zeta}) we have 
 \begin{equation}
 E^{div }=\fr{a_2}{32 \pi^2}(\fr{1}{s}-2+ln\fr{4 \mu^2}{m^2}), 
 \label{e_div_def}
 \end{equation}
where $a_2$ is the only non-zero heat kernel coefficient $a_n,\ 0 <
n\le2$, contributing to the divergent part of vacuum energy.

For the renormalized ground state energy the asymptotical dependence
on powers of $m$ at $m\rightarrow \infty$ follows from
Eq.(\ref{heat_kt_sum}) to have the form
 \begin{equation}
 E_0^{ren}\stackrel{\sim }{_{m\rightarrow\infty}} \sum\limits_{n > 2} \fr{e_n}{m^{2n}}  
 \end{equation}
with some coefficients $e_n$.

In accordance with the interpretation of $ E^{ren}$ it must vanish in
the limit of $m\rightarrow \infty$ since it is the energy of vacuum
fluctuations.  Through the subtraction of terms containing all
non-negative powers of $m^2$ (which are the terms of heat kernel
expansion up to $a_2$) the condition (\ref{norm_cond}) is satisfied
automatically.
 
Some comments on the subtraction scheme are in order.  This scheme is
motivated by the physical assumption that the quantum fluctuations
should vanish if the mass of the fluctuating field becomes large. The
scheme had been used in a number of Casimir energy calculations.  In
\cite{Bordag:2002dg} it had been shown to be equivalent to the so
called ``no tadpole'' normalization condition which is common in field
theory.  It should be noticed that this scheme does not apply to
massless fluctuating fields, for a discussion of this point see
\cite{Bordag:1998vs}.

So $E^{ren}$ as given by Eq.(\ref{e0-ediv})is now finite at
$s\rightarrow0, $ but it is not possible to use this expression
immediately since the integral in this limit does not exist; i.e. one
cannot carry out the analytical continuation to $s=0$.  In order to
get a representation where this can be done one can make a trick
\cite{bordag}, namely add and subtract a part $\ln f_l^{asym}$ of the
uniform asymptotical expansion for $\ln f_l$ in (\ref{e0_of_lnf})
containing a minimal number of terms to provide the convergence of the
remaining part after the subtraction.  Thus we can separate the
resulting expression into four terms
\begin{equation}
E^{ren}=(E_0 - E^{asym}) + (E^{as}-E^{div }) 
\label{eren_4parts}
\end{equation}
where 
\begin{equation}
 E^{asym}=2C_s\sum\limits_{l}^{}\int\limits_{m}^{\infty}dk(k^2-m^2)^{1-s}
\fr{\pd}{\pd k} \ln f_l^{asym}(ik)
\label{e0_of_lnfas}
\end{equation}
and redefine two terms in brackets as
\begin{eqnarray}
E^{ren}=E^f + E^{as}\nn\\
E^f=E_0 - E^{asym}\\
E^{as}=E^{asym}-E^{div },
\label{eren_ef_eas_ediv}
\end{eqnarray}
thus $E^f$ is defined to be  
\begin{equation}
E^f=\fr{1}{2\pi}\sum\limits_{l=-\infty}^{\infty}\int\limits_{m}^{\infty}dk\
(k^2-m^2)\fr{\pd}{\pd k}[ ln f_l (ik)-\ln f^{as}_l(ik)]
\label{ef_pure}
\end{equation}  
The ``finite'' part $E^f$ of the renormalized vacuum energy is not only well defined
at $s \rightarrow 0 $, but also provides a representation well suited for
numerical analysis. The analytical continuation for $E^{as}$ at this
limit will be constructed below.

\section{ Solution of the Dirac equation } 
We consider a spinor quantum field $\psi$ in the background of the
classical magnetic flux. We start with the Dirac equation for this
field 
 \begin{equation}
 \left\{i \gamma^{\mu} \fr{\pd}{\pd x^{\mu} }-m-e \gamma^{\mu}
 A_{\mu}\right\}\psi =0 
\label{dirac_eq}
\end{equation}
with the electromagnetic potential (\ref{potential_A}).
The $\psi$ in (\ref{dirac_eq}) is a spinor quantum field of mass $m$ and charge
$e$, interacting with $\vec{A}$ (\ref{potential_A}). 
The gamma matrices in our representation are chosen to be the same as in
\cite{borvor} 
\begin{equation}
\gamma^0=\left\{\begin{array}{cc} \sigma_3 & 0\\0 & -\sigma_3
 \end{array}\right\} \ \ \ \ \ \gamma^1=\left\{\begin{array}{cc} i\sigma_2 & 0\\0 & -i\sigma_2 \end{array}\right\} \\
\\
\gamma^2=\left\{\begin{array}{cc} -i\sigma_1 & 0\\0 & i\sigma_1
 \end{array}\right\} \ \ \ \ \ \gamma^3=\left\{\begin{array}{cc} 0 &
 1\\-1 & 0 \end{array}\right\}.
\end{equation}
Now we follow the standard procedure and separate the variables.
Using the ansatz
\begin{equation}
\phi(r, \varphi, z) = e^{-ip_0x^0} e^{-ip_z z}\Psi(r, \vp)\\
\end{equation}
 we obtain the equation for the 4- component spinor
$ \Psi=\left[ \begin{array}{c} \phi \\ \chi \end{array}\right]$ 
 \begin{equation}
\left\{\begin{array}{cc} p_0+\hat{L}-m\sigma_3 & p_3\sigma_3\\ \\p_3 \sigma_3
    & p_0+\hat{L}+m\sigma_3 
 \end{array}\right\}\Psi=0, 
\end{equation}
where $\hat{L}=i\sum\limits_{i=1}^{2}\sigma_i(\pd_i+ieA_i)$, 
$\phi, \chi $ are the two-component spinors.

Respecting the translational invariance of the system it is sufficient
to solve the equations only for $p_3=0$. Then one of the two decoupled
equations $(\phi, \chi)$ reads 
\begin{equation}
\left\{\begin{array}{cc} p_0-m & \fr{\pd}{\pd r}-\fr{l-\delta
 a(r)}{r}\\ \\-\fr{\pd}{\pd r}-\fr{l+1-\delta a(r)}{r} & p_0+m 
 \end{array}\right\}\Phi(r)=0
\label{dirac_of_r}
\end{equation}
(here the standard ansatz
\begin{equation}
\Phi =\left[ \begin{array}{c} i\psi^u(r) e^{-i(l+1)\vp} \\ \\ \psi^l(r)
 e^{-il\vp} \end{array}\right]
\label{ansatz_phi}
\end{equation}
has been used, $l $ is the orbital quantum number), the radial part denoted as
\begin{equation}
\Phi(r)=\left\{\begin{array}{c} \psi^{u}(r)\\ \\ \psi^{l}(r)  
\end{array}\right\} 
\end{equation}
and
\begin{equation}
\delta=\fr{e\Phi}{2\pi}.
\label{delta}
\end{equation}

In order to use later the symmetry properties of the Jost
function we redefine the parameter $l$ in (\ref{ansatz_phi}) (orbital number)
as $\nu$ according to
\begin{equation}
\nu=\left\{ \begin{array}{c} l+\half\ for\ l=0, 1, 2, . . . \\ \\
 -l-\half\ for\ l=-1, -2, . . . \end{array}\right.
\end{equation}

For the given construction of potential (\ref{a_r1})
we get three equations and three types of solutions for 3 areas of space respectively\\
Domain I: $ r<R_1 $. The free wave equation (the magnetic flux is zero), 
the solutions are the Bessel functions $J_{\nu\pm\half}(kr)$ 
\begin{eqnarray}
\Phi_{I}^{-}(r)= \left[\begin{array}{c} \psi^{u-}_{I}(r)\nn\\ \nn\\\psi^{l-}_{I}(r) \end{array} \right]
 ={ \fr{1}{\sqrt{p_0-m}}\left[
\begin{array}{c}-\sqrt{p_0+m}J_{\nu-\half}(k r)\nn\\ \nn\\ \sqrt{p_0-m}J_{\nu+\half}(k r)\end{array} \right]},
\nn\\
\Phi_{I}^{+}(r)= \left[\begin{array}{c} \psi^{u+}_{I}(r)\nn\\ \nn\\\psi^{l+}_{I}(r) \end{array} \right]
={\fr{1}{\sqrt{p_0-m}}\left[
\begin{array}{c}\sqrt{p_0+m}J_{\nu+\half}(k r)\nn\\ \nn\\
 \sqrt{p_0-m}J_{\nu-\half}(k r)\end{array} \right]}
\end{eqnarray}
\label{domain_I}
\hspace*{3cm}$ p_0=\sqrt{m^2-k^2} $\\
 This solution of equation (\ref{dirac_of_r}) in the domain I is chosen
 to be the so called regular solution which is defined as to coincide for
 $r\rightarrow0$ with the free solution.\\ 
Domain II: $R_1<r<R_2$ The equation with a homogeneous magnetic field\\
has the solution
\begin{eqnarray}
\Phi_{II}^{-}(r)=
C_r^{-}\left[\begin{array}{c} \psi^{u-}_{II. r}(r)\nn\\
 \nn\\\psi^{l-}_{II. r}(r) \end{array} \right] +
C_i^{-}\left[\begin{array}{c} \psi^{u-}_{II. i}(r)\nn\\
 \nn\\\psi^{l-}_{II. i}(r) \end{array} \right]=\nn\\ \nn\\
\nn\\
C_r^{-} \left[
 \begin{array}{c}
 \fr{(p_0+m)}{2(\tilde{\alpha}+1)}r^{\tilde{\alpha}+1} e^{-\fr{\beta
 r^2}{4} } {_1}F_1(1-\fr{k^2}{2 \beta}, \tilde{\alpha}+2 ;\fr{\beta r^2}{2})
 \nn\\ \nn\\ \hfill r^{\tilde{\alpha}} e^{-\fr{\beta
 r^2}{4}} {_1}F_1(-\fr{k^2}{2 \beta}, \tilde{\alpha}+1 ;\fr{\beta r^2}{2} )
 \end{array} 
 \right]\nn\\
 \nn\\ \\
 + C_i^{-} \left[
 \begin{array}{c}
 \fr{1}{p_0-m}\fr{2 \tilde{\alpha}}{r}(\fr{\beta
 r^2}{2})^{-\tilde{\alpha}} e^{-\fr{\beta r^2}{4}}{_1}F_1(-\fr{k^2}{2
 \beta}-\tilde{\alpha}, -\tilde{\alpha} ;\fr{\beta r^2}{2})\nn\\
 \nn\\ (\fr{\beta
 r^2}{2})^{-\tilde{\alpha}} e^{-\fr{\beta r^2}{4}} {_1}F_1(-\fr{k^2}{2
 \beta}-\tilde{\alpha}, 1-\tilde{\alpha} ;\fr{\beta r^2}{2})
 \end{array} 
 \right]. 
\end{eqnarray}
$ \Phi_{II}^{+}(r) $ is the same, but $\tilde{\alpha}$ replaced by $\alpha$\\
here: $\alpha=\nu-\half+\fr{\delta}{\kappa}$ for $ l\ge 0$\\
\hspace*{2cm}$\tilde{\alpha}=\fr{\delta}{\kappa}-\nu-\half $ for $ l< 0$\\
\hspace*{2cm}$\beta=\fr{2 \delta}{\kappa R_1^2}$\\
${_1}F_1$ is a confluent hypergeometric function \cite{abrsteg}, 
\cite{gradst}. 

The coefficients $C_i , C_r$ are some constants that will be irrelevant for expressing the Jost function. The indices u and l
denote ``upper'' and ``lower'' components of spinor respectively. The
lower index ``i'' or ``r'' corresponds to ``regular'' or ``irregular''
part of solutions dependent on the behaviour of the function if
continued to $r=0$. If the external background vanishes ($ \delta
\rightarrow 0 $), contributions of irregular parts to the solution
disappear.\\
Domain III: $r>R_2$ The free wave equation outside of the magnetic flux has
the solutions
\begin{eqnarray}
 \Phi_{III}^{-}(r)&=&
\half \bar{f}^{-}_{\nu}(k)  \fr{1}{\sqrt{p_0-m}}\left[\begin{array}{c} \psi^{u-}_{III. r}(r)\nn\\ \nn\\\psi^{l-}_{III. r}(r) \end{array} \right]
+\half f^{-}_{\nu}(k)\fr{1}{\sqrt{p_0-m}}\left[\begin{array}{c} \psi^{u-}_{III. i}(r)\nn\\ \nn\\\psi^{l-}_{III. i}(r) \end{array} \right]
 \nn\\
 &=&\half \bar{f}^{-}_{\nu}(k)  \left[
\begin{array}{r}-\eta H^{(1)}_{\nu-\half+\delta}(k r)\nn\\ \nn\\ H^{(1)}_{\nu+\half+\delta}(k
 r)\end{array} \right]+\half f^{-}_{\nu}(k) \left[
\begin{array}{r}-\eta H^{(2)}_{\nu-\half+\delta}(k r)\nn\\ \nn\\  H^{(2)}_{\nu+\half+\delta}(k
 r)\end{array} \right],\nn\\
\nn
\end{eqnarray}
\\
\begin{eqnarray}
 \Phi_{III}^{+}(r)&=& \half \bar{f}^{+}_{\nu}(k)\fr{1}{\sqrt{p_0-m}} \left[\begin{array}{c} \psi^{u+}_{III. r}(r)\nn\\
 \nn\\\psi^{l+}_{III. r}(r) \end{array} \right]
+\half f^{+}_{\nu}(k)\fr{1}{\sqrt{p_0-m}} \left[\begin{array}{c} \psi^{u+}_{III. i}(r)\nn\\ \nn\\\psi^{l+}_{III. i}(r) \end{array} \right]
 \nn\\
 &=& \half \bar{f}^{+}_{\nu}(k) \left[
\begin{array}{r}\eta H^{(1)}_{\nu+\half-\delta}(k r)\nn\\ \nn\\ H^{(1)}_{\nu-\half-\delta}(k
 r)\end{array} \right]+\half f^{+}_{\nu}(k) \left[
\begin{array}{r}\eta H^{(2)}_{\nu+\half-\delta}(k r)\nn\\ \nn\\ H^{(2)}_{\nu-\half-\delta}(k
 r)\end{array} \right],\nn\\ \nn\\
\eta &=&\fr{\sqrt{p_0+m}}{\sqrt{p_0-m}}\nn
\end{eqnarray}
and $f_{\nu}(k)$ is the desired Jost function in accordance to the
definition ( \ref{jost_lk}). The functions $ f_l(-k)$ and $
\bar{f}_l(k)$ are conjugated to each other because of the choice of
the solution in the domain I to be regular (\ref{domain_I}).

\section{The exact and asymptotic Jost function for the background}
To obtain a Jost function we need to impose certain matching conditions
for the spinor wave at the boundaries between I-II and II-III
domains.
It is known that for a continuous potential $a(r)$ as we consider here, the
spinor wave function must be continuous, hence for its components it holds
\begin{eqnarray}
\Phi_{II}^{\mp}(r) =\Phi_{I}^{\mp}(r)\Big|_{r=R_1}, \nn\\
\Phi_{III}^{\mp}(r) =\Phi_{II}^{\mp}(r)\Big|_{r=R_2} 
\end{eqnarray}
Resolving these equations for $f_{\nu}^{\pm}$ we obtain, that \\
\begin{eqnarray}
f_{\nu}^{\pm} (k)&=& 2\fr{ A + B }{ W_{III} W_{II} }\nn\\
 A &=& (\psi_{I}^{u\pm}\psi_{II. i}^{l\pm}-\psi_{I}^{l\pm}\psi_{II. i}^{u\pm})\Big|_{R_1}(\psi_{II. r}^{u\pm}\psi_{III. r}^{l\pm}-\psi_{II. r}^{l\pm}\psi_{III. r}^{u\pm})\Big|_{R_2},\nn\\
B &=& (\psi_{I}^{l\pm}\psi_{II. r}^{u\pm}-
 \psi_{I}^{u\pm}\psi_{II. r}^{u\pm})\Big|_{R_1}(\psi_{II. i}^{u\pm}\psi_{III. r}^{l\pm}-\psi_{II. i}^{l\pm}\psi_{III. r}^{u\pm})\Big|_{R_2}
\end{eqnarray}

The denominator of this expression can be written using the Wronskians
of hypergeometric and Bessel functions as follows
\begin{eqnarray}
W_{III}=(\psi_{III. i}^{u}\psi_{III. r}^{l}-\psi_{III. i}^{l}\psi_{III. r}^{u})\Big|_{R_2}=
(\fr{p_0+m}{p_0-m})^{\half} \fr{4i}{\pi k R_2}\nn\\
\\W_{II}=(\psi_{II. r}^{u}\psi_{II. i}^{l}-\psi_{II. r}^{l}\psi_{II. i}^{u})\Big|_{R_1}=
-\fr{\alpha\beta}{p_0-m}(\fr{\beta R_1}{2})^{-\alpha-1}
\end{eqnarray} 

To calculate $E_0$ we need this function with imaginary argument
(\ref{e0_of_lnf}) so we need to replace $k$ by $ik$. As a result we obtain
the new expression for $f(ik)$ that contains now modified Bessel functions $I_{\nu\pm \half}$ instead
of $J_{\nu\pm \half}$ and modified Bessel functions $K_{\nu\pm \half\mp \delta}$
instead of $H^{(1, 2)}_{\nu\pm \half\mp \delta}$.  
\begin{eqnarray}
 f^{+}_\nu(ik) &=& \fr{\pi k R_1 R_2 e^{-\delta /\kappa+\fr{i\pi\delta}{2}}}
 {\half- \tilde{\nu}} \times \nn \\
&& \left\{ \left[ \fr{kR_1}{2\tilde{\nu}+1}\  
{_1}F_1(1+\fr{k^2}{2\beta},\ \tilde{\nu}+\fr{3}{2};\ \fr{\beta R_1^2}{2})
\ I_{\nu-\half}(k
 R_1)-\right.\right.\nn\\&&\left.\hspace*{4cm}{_1}F_1(\fr{k^2}{2\beta},\
 \tilde{\nu}+\half;\ \fr{\beta R_1^2}{2})
\ I_{\nu+\half}(kR_1) \right]\times\nn\\ 
&&\left[\fr{2\tilde{\nu}+1}{R_2}\
 {_1}F_1(\half-\tilde{\nu}+\fr{k^2}{2\beta},\ \half-\tilde{\nu};\ \fr{\beta R_2^2}{2})\ K_{\nu-\half-\delta}
 (kR_2)-\right.\nn\\ 
&&\left.\hspace*{4cm} k\ {_1}F_1(\half-\tilde{\nu}+\fr{k^2}{2\beta},\
 \fr{3}{2}-\tilde{\nu};\ \fr{\beta R_2^2}{2})\ K_{\nu+\half-\delta}(kR_2)\right]- \nn\\
&&\left[k\
 {_1}F_1(\half-\tilde{\nu}+\fr{k^2}{2\beta},\ \fr{3}{2}-\tilde{\nu};\ \fr{\beta
 R_1^2}{2})\ I_{\nu+\half}(kR_1)+\right.\nn\\
&&\left.\hspace*{4cm}\fr{2\tilde{\nu}-1}{R_1}\
 {_1}F_1(\half-\tilde{\nu}+\fr{k^2}{2\beta},\ \half-\tilde{\nu};\ \fr{\beta
 R_1^2}{2})\ I_{\nu-\half}(kR_1)\right]\times\nn\\ 
&&\left[\fr{kR_2}{2\tilde{\nu}+1}\
 {_1}F_1(1+\fr{k^2}{2\beta},\ \tilde{\nu}+\fr{3}{2};\ \fr{\beta R_2^2}{2})\
 K_{\nu-\half-\delta}(kR_2)+\right.\nn\\&&\left.\left.\hspace*{4cm}{_1}F_1(\fr{k^2}{2\beta},\
 \tilde{\nu}+\half;\ \fr{\beta
 R_2^2}{2})\ K_{\nu+\half-\delta}(kR_2)\right]\right\},\nn
\end{eqnarray}
where $\tilde{\nu}= \nu + \fr{\delta R_1^2}{R_1^2-R_2^2} $, and

\begin{eqnarray}
 f^{-}_\nu(ik) &=& \fr{\pi k R_1 R_2 e^{-\delta /\kappa-\fr{i\pi\delta}{2}}}
 {\half+ \bar{\nu}} \times \\
&& \left\{ \left[ \fr{kR_1}{2\bar{\nu}-1} 
\ {_1}F_1(\bar{\nu}+\half+\fr{k^2}{2\beta},\ \bar{\nu}+\half;\ \fr{\beta R_1^2}{2})
\ I_{\nu+\half}(rR_1)-\right.\right.\nn\\
&&\left.\hspace*{4cm}{_1}F_1(\bar{\nu}+\half+\fr{k^2}{2\beta},\
 \bar{\nu}+\fr{3}{2};\ \fr{\beta R_1^2}{2})
\ I_{\nu-\half}(kR_1) \right]\times\nn\\ 
&&\left[\fr{2\bar{\nu}-1}{R_2}\  
  {_1}F_1(1+\fr{k^2}{2\beta},\ \fr{3}{2}-\bar{\nu};\ \fr{\beta R_2^2}{2})
  \ K_{\nu+\half+\delta}(kR_2)-\right.\nn\\
&&\left.\hspace{4cm}
 k\ {_1}F_1( \fr{k^2}{2\beta},\ \half-\bar{\nu};\ \fr{\beta R_2^2}{2})
\ K_{\nu-\half+\delta}(kR_2)\right]- \nn\\
&&\left[k\ {_1}F_1(\fr{k^2}{2\beta},\ \half-\bar{\nu};\ \fr{\beta R_1^2}{2})
\ I_{\nu-\half}(kR_1)+\right.\nn\\
&&\left.\hspace*{4cm}\fr{2\bar{\nu}+1}{R_1}
  \ {_1}F_1(1+\fr{k^2}{2\beta},\ \fr{3}{2}-\bar{\nu};\ \fr{\beta R_1^2}{2})
  \ I_{\nu+\half}(kR_1)\right]\times\nn\\ 
&&\left[\fr{kR_2}{2\bar{\nu}-1}
  \ {_1}F_1(\bar{\nu}+\half+\fr{k^2}{2\beta},\ \bar{\nu}+\half;\ \fr{\beta R_2^2}{2})
  \ K_{\nu+\half+\delta}(kR_2)+\right.\nn\\
&&\left.\left.\hspace*{4cm}{_1}F_1(\bar{\nu}+\half+\fr{k^2}{2\beta},\
 \bar{\nu}+\fr{3}{2};\ \fr{\beta R_2^2}{2})
 \ K_{\nu-\half+\delta}(kR_2)\right]\right\},\nn
\label{f_v_of_ik}
\end{eqnarray}
where $\bar{\nu}= \nu - \fr{\delta R_1^2}{R_1^2-R_2^2} $, and

Now we need to obtain the asymptotic Jost function and asymptotic part
of energy.  We use the representation of the asymptotic Jost function
$\ln f_{as}\unboldmath(ik)$ in the following form (see App.)
\begin{equation}
ln \boldmath f_{as}\unboldmath(ik)=\sum\limits_{n=1}^3\sum\limits_{j=1}^9
\int\limits_{0}^{\infty} \fr{dr}{r}X_{nj}\fr{t^j}{\nu^n} 
\label{lnfas}
\end{equation}
where $t=(1+[\fr{kr}{v}]^2)^{-\half}$ and $X_{nj}$ (given
explicitly in App.A) are represented in terms of $ r, \delta, a(r) $ and
their derivatives.

This expansion had been obtained in \cite{borkir2} by iterations of
Lippmann-Schwinger equation up to the order $\nu^{-3}$. In general it
is possible to obtain higher orders using this formalism. However as
the calculations \cite{borkir2} showed,the complication of the
involved expressions increases very fast. It is remarkable that this
expression does not contain a term with power $\nu^{-2}$. In the
finite part of the energy the corresponding term is canceled in the
sum of terms corresponding to positive and negative orbital momenta
$\nu$ as well. The absence of the power $\nu^{-2}$ is a succession of
the zero heat kernel coefficient $a_{3/2}$.  Also it is a non-trivial
fact that both, the
fourth power of the magnetic flux $\delta$ and the second one is present.

It can be checked numerically that $ln f^{as}$ (\ref{lnfas}) is indeed
the uniform asymptotic expansion of logarithm of (\ref{f_v_of_ik}) for
$k,\nu$- large, $k/ \nu=z$-fixed.
 
\section{The finite and asymptotic parts of vacuum energy}
The expression (\ref{lnfas}) for the uniform asymptotic of Jost function, 
substituted into the expression for $E^{as}$ (\ref{eren_ef_eas_ediv}) yields
\begin{equation}
E^{as}=2C_s\sum\limits_{v=1/2, 3/2, . . . }^{}\int\limits_{m}^{\infty}dk(k^2-m^2)^{1-s}
\fr{\pd}{\pd k}\int\limits_{0}^{\infty}\fr{dr}{r}\sum\limits_{n=1}^3\sum\limits_{j=1}^9
\int\limits_{0}^{\infty}X_{nj}\fr{t^j}{\nu^n} - E^{div}\\
\label{eas-ediv}
\end{equation}
with the constant $C_s =\fr{1}{2\pi}[1+s(-1+2ln2\mu)]$. $E^{div}$ as defined in (\ref{e_div_def}) is
\begin{equation}
  E^{div}=\fr{\delta^2}{6\pi(R_2^2-R_1^2)}(\fr{1}{s}-2+\ln \fr{4\mu^2}{m^2})
\end{equation}
with the coefficient $a_2$ according to (\ref{coeff_a_2}) has been used in the
form 
\begin{equation}
a_2=\fr{2}{3}F^2_{\mu\nu}
=\fr{8 \pi}{3}\delta^2\int\limits_{0}^{\infty}dr r h(r)^2 = \fr{16 \pi\delta^2}{3}\fr{1}{R_2^2-R_1^2}  
\end{equation}

Then the sum over $\nu$ can be now transformed into two integrals by
means of Abel-Plana formula (\ref{abelplan}). The first one cancels
the $E^{div}$ exactly.  Then we integrate the second one over $k$
using the identities (\ref{ident1}, \ref{ident2}). It gives the form
\begin{equation}\label{Eas}
 \hspace*{1cm}E^{as}=-\fr{1}{\pi}m^2\sum\limits_{n=1}^{3}\sum\limits_{j=n}^{3n}\int\limits_{0}^{\infty}\fr{dr}{r}X_{n, j}\Sigma_{n\
 j}(rm)
\end{equation}
with
\begin{eqnarray}
\hspace*{1cm}&&\Sigma_{n\ j}(x) =\fr{
  \Gamma(s+j/2-1)}{\Gamma(j/2)}\fr{-i}{x^j} \\&\times&
\int\limits_{0}^{\infty}\fr{d\nu}{1+exp(2 \pi\nu)} \left( \fr{(i\nu)^{j-n}}{[1+(i\nu/x)^2]^{s+j/2-1}}- \fr{(-i\nu)^{j-n}}{[1+(-i\nu/x)^2]^{s+j/2-1}} \right)\nn
\end{eqnarray}

In order to obtain an analytical continuation in $s=0$ of each term of
the sum we integrate it over $\nu$ by parts several times till the
divergency at $\nu=0$ through the power $s+j/2-1$ abrogates, and after
that we can perform the integration over $r$. Further we integrate by
parts resulting in the relations $raa' \rightarrow -\half a^2 r\pd_r$
and $r^2aa'' \rightarrow -a'^2+\half a^2 r\pd^2_r r $ (that hold
because of the continuity of the potential $a(r)$), and we obtain
finally the form
\begin{equation}
E^{as}= \fr{-4}{\pi}\int\limits_{0}^{\infty}\fr{dr}{r^3}[\delta^2 a(r)^2
g_1(rm)-\delta^2 r^2 a'(r)^2 g_2(rm)+\delta^4 a(r)^4 g_3(rm)]\\
\label{eas_of_g}
\end{equation}
with $ \ \ \ \ g_i(x)=\int\limits_{x}^{\infty}d \nu
\sqrt{\nu^2-x^2}f_i(\nu)$ 
that will be calculated numerically (see the next section), the functions $
f_i(\nu)$ are shown explicitly in (\ref{f_i_of_v}).

The finite part of the ground state energy $E^f\ $ can be finally represented as follows
\begin{equation}
E^f=\fr{1}{2\pi}\sum\limits_{\nu=1/2, 3/2,
  . . . }^{}\int\limits_{m}^{\infty}dk\ (k^2-m^2)\fr{\pd}{\pd k}[ ln f_{\nu}^{+}(ik)+ln f_{\nu}^{-}(ik)-2 \ln
f^{as}(ik)]
\label{ef}
 \end{equation}
In order to use this form for numerical evaluations we integrate by
parts and get
\begin{equation}\label{ef1}
E^f=-\fr{1}{\pi}\sum\limits_{\nu=1/2, 3/2,
  . . . }^{}\int\limits_{m}^{\infty}dk\ k\ ln f_{\nu}^{sub}(ik)
\end{equation}
(the logarithmic expression in square brackets in (\ref{ef}) denoted as $ ln
f^{sub} $).

\section{Higher orders of the uniform asymptotic expansion of the Jost
  function and the heat kernel coefficient $a_{5/2}$}
The background considered in this paper has singular surfaces where
the magnetic field jumps. The heat kernel expansion for the case of
singularities concentrated at surfaces has been considered in
[\cite{borvass}, \cite{moss}, \cite{gkv}, \cite{gkv2}].  Although the
general analysis of \cite{gkv} is valid for our background, an
explicit expression for $a_{5/2}$ has not been calculated yet.
 
We can use our obtained Jost function (\ref{f_v_of_ik}) to calculate
 the coefficient $a_{5/2}$ in the heat kernel expansion
 (\ref{heat_kt_sum}).
 
Suppose we have obtained the value of $E_0(s)$ (\ref{e0_of_lnf})in the
 point $s=-1/2$. It follows from (\ref{e0_zeta} - \ref{heat_kt_sum})
 that
 \begin{equation}
 E_0(s)\sim - \fr{\mu^{2s}}{2(4\pi)^{3/2} } \fr{m^{4-2s}}{\Gamma(s-\half)}
 \sum\limits_{n=0, \half, 1, . . . }\fr{a_n}{m^{2n}} \Gamma(s-2+n)
 \end{equation}
and at the limit of $s \rightarrow -\half$ only the term containing
 $a_{5/2}$ remains to be nonzero in the sum.
\begin{equation}
 E_0(\half)= - \fr{\mu^{2s}}{2(4\pi)^{3/2} } a_{5/2}
\end{equation}

From the other hand we have (\ref{e0_of_lnf}, \ref{e0_of_lnfas})
\begin{equation} 
 E_0(s)= C_s \sum\limits_{\nu=1/2, 3/2, . . . } \int\limits_m^\infty
 dk(k^2-m^2)^{1-s} \fr{\pd}{\pd k} f_\nu(ik)=C_s h(s).
\end{equation} 
Here we substitute the exact Jost function by its uniform asymptotic
represented in the form
\begin{equation} 
 f^{ua}_\nu(ik)=\sum\limits_{n=1,3,4,5...} \fr{h_n(t_1, t_2)}{\nu^n},
 \label{unif_as_f}
\end{equation}
where the coefficients $h_n$ are functions of
$t_1=(1+(\fr{kR_1}{\nu})^2)^{-1/2}$,
$t_2=(1+(\fr{kR_2}{\nu})^2)^{-1/2}$, the power $\nu^{-2}$ is absent,
as noticed above (\ref{lnfas}),
\begin{equation} 
C_s= -\fr{\mu^{2s}\Gamma(s-1)}{4 \sqrt{\pi}\Gamma(s-1/2)}\fr{-4 \sin(\pi s)}{\pi}
\end{equation}
At the limit $s \rightarrow -\half$ it yields 
\begin{equation} 
 E_0(-\half)=\fr{4}{3\pi} Res_{s \rightarrow -\half} h(s) 
\end{equation}
and therefore we obtain for $a_{5/2}$
\begin{equation}
 a_{5/2}=-\fr{64\sqrt{\pi}}{3} Res_{s \rightarrow -\half} h(s).
\end{equation}

To obtain the explicit form of $Res_{s \rightarrow -\half} h(s)$ we
 use the uniform asymptotic expansion of the Jost function
 (\ref{unif_as_h}).  The terms $h_n(t_1,t_2)$ in \ref{unif_as_f} can
 be obtained either by iterations of Lippmann-Schwinger equation (see
 \cite{borkir2}, \cite{borkir1}) or by using the explicit form of the
 Jost function as well. All the further terms up from $h_4$ are
 produced from the explicit form of the Jost function
 (\ref{f_v_of_ik}) because of complication of the first way for higher
 orders $n$ (see the remark to \ref{lnfas} in Sec.6).  Namely, we
 obtain several higher orders $1/\nu$ of uniform asymptotic expansion
 for special functions $I_{\nu},K_{\nu},_{1}F_1$ which the exact Jost
 function (\ref{f_v_of_ik}) consists of (it can be done starting with
 the explicit form for two first orders and executing the recursive
 algorithm several times\cite{abrsteg}), then after substitution of
 each of functions $I ,K, _{1}F_1$ by its corresponding uniform
 asymptotic expansion and separation of powers of $\nu$ we arrive at
 the form (\ref{unif_as_h}). The coefficients $h_n(t_1, t_2),
 n=1,...,4$ are given in the Appendix (\ref{unif_as_h}).
 
 If the function $h_n(t_1, t_2)$ is a polynomial over $t_1, t_2$, so
 we can consider some term $t^j , (t=t_i, i=1, 2 ) $ of it.  Notice,
 that for $h_1(t_1,t_2),h_3(t_1,t_2)$ it is not the case, but we can
 treat the terms of kind $\fr{1}{1+t}$ and $\fr{1}{(1+t)^k}$,
 $k$-integer, as an infinite sum of powers $t$.
 
 Apart from the construction of $t= t_1, t_2$ (\ref{unif_as_f}) they
 are strictly positive and less than 1, therefore the series $
 \fr{1}{1+t}=\sum\limits_{i=0}^{\infty}t^i$ converges regular and
 uniform and respecting that we have only finite integer powers $k$ so
 that $ \left[ \fr{1}{1+t}\right]^k=
 \left(\sum\limits_{i=0}^{\infty}t^i\right)^k $ converges as well, the
 sum over $i$ can be interchanged with the one over $\nu$
 (Princeheim's Theorem) thus the following procedure is valid.
 Performing the sum
\begin{equation}
 h(s)=\sum\limits_{\nu=1/2, 3/2, . . . } \int\limits_m^\infty
 dk(k^2-m^2)^{1-s} \fr{\pd}{\pd k} \sum\limits_{n=1, 3, 4, 5} h_{n,j}(R_1,R_2) \fr{t^j}{\nu^n}
\end{equation}
by meaning of (\ref{abelplan}) we obtain the sum of two parts
\begin{eqnarray}
h(s)&=&\int\limits_0^\infty d\nu\int\limits_m^\infty dk(k^2-m^2)^{1-s}\fr{\pd}{\pd
 k}\sum\limits_n h_{n,j}(R_1,R_2)\fr{t^j}{\nu^n}+ \\\nn
&&\int\limits_0^\infty \fr{d\nu}{e^{2\pi\nu}+1}\fr{1}{i}\left[ \int\limits_m^\infty
 dk(k^2-m^2)^{1-s}\fr{\pd}{\pd k}\sum\limits_n h_{n,j}(R_1,R_2)\fr{t^j}{\nu^n}
\right]_{\nu=i\nu}^{\nu=-i\nu},
\label{two_parts}
\end{eqnarray}
where $t=t_{1, 2}$. The first summand of (\ref{two_parts}) gives for each power $j$ of $t=\{ t_1, t_2 \}$ 
(using the (\ref{ident1}))
\begin{equation}
 -\fr{m^3}{2}
  \fr{\Gamma(2-s)\Gamma(\fr{1+j-n}{2})\Gamma(s+\fr{n-3}{2})}{(Rm)^{n-1}\Gamma(j/2)}
\label{first_part}  
\end{equation}
(where R denotes $R_1$ and $R_2$ respectively)
at the limit $s \rightarrow -\half$ only the terms corresponding to $n=2$
and $n=4$ (which will be calculated below) have a pole.

For the second part of the Abel-Plana formula (\ref{two_parts}) we have
\begin{eqnarray}
&&\fr{1}{i}\int\limits_0^\infty\fr{d\nu}{1+e^{2\pi\nu}}
\fr{\Gamma(2-s)\Gamma(s+j/2-1)}{jR^{4-2s}\Gamma(j/2)} \\
&\times&\left[(i\nu)^{2+j-n}[(mR)^2+(i\nu)^2]^{1-j/2-s}-(-i\nu)^{2+j-n}[(mR)^2+(-i\nu)^2]^{1-j/2-s}\right]\nn
\label{2_part}
\end{eqnarray}

It can be seen using the Taylor expansions at $s \rightarrow -\half$
of Gamma functions entering the (\ref{2_part}), that only the terms
containing -1, 1, and 3 powers of $t_i$ can contribute to the residuum
at $s \rightarrow -\half$. But the expression in square brackets in
(\ref{2_part}) can be performed as
\begin{equation}
 -\nu^2(\nu^2-(mr)^2)[(i\nu)^{j-n}e^{i\pi (1-j/2-s)}-(-i\nu)^{j-n}e^{-i\pi (1-j/2-s)}]
\end{equation}
thus for $j-n$ even the expression in square brackets produces (dropping the
non sufficient coefficient $\pm 2$ or $\pm 2i$):\\
\hspace*{2cm}for $j=-1, 1 :\ \ \ \ \sin \pi(1-j/2-s)=\cos \pi s $\\
\hspace*{2cm}for $j=3 :\ \ \ \ \sin \pi(1-j/2-s)= -\cos \pi s $\\
\\
and for $j-n$ odd respectively\\
\hspace*{2cm}for $j=\pm 1 :\ \ \ \ \cos \pi(1-j/2-s)=\pm\sin \pi s $\\
\hspace*{2cm}for $j=3 :\ \ \ \ \cos \pi(1-j/2-s)= -\sin \pi s $\\
\\
But for $s \rightarrow -\half$ these functions behave as\\
\hspace*{2cm}$\sin \pi s \sim -1+\fr{\pi^2}{2}(\half+s)^2, \\ \cos \pi
s \sim \pi(\half+s)\\ $ and it means that only the contribution of
terms with odd powers $j-n$ of $i\nu$ could survive and these
corresponds for the possible values of $j$ to $\nu^{-2}, \nu^{-4},
\nu^{-6}, . . . $. In fact the coefficient $h_2(s)$ at $\nu^{-2}$ is
zero, and all other possible terms up from $\nu^{-4}$ does not contain
any powers of $t_i$ lower than 4; one can see it for example in the
explicit form of the uniform asymptotical expansion of special
functions entering the $\ln f_v(ik)$ (\ref{f_v_of_ik}). Therefore the
second summand of (\ref{two_parts}) does not produce any contribution to
the $Res_{s \rightarrow -\half} h(s)$ Thus we have that only the
contribution from the first summand of (\ref{two_parts}) remains, and
since the term of $n=2$ is zero the searched residuum resulting from
the term of $n=4$ is:
\begin{equation}
 Res_{s \rightarrow -\half} h(s)=-\half.\fr{3 \sqrt{\pi}}{4}
 \fr{\delta^2}{4(R_1^2-R_2 ^2)^2} (R_1+R_2)\sum\limits_{j=4, 6}\fr{\Gamma(\fr{1+j-n}{2})}{j/2},
 \end{equation} 
 where the (\ref{first_part}) and the explicit form of $ h_4(t_1, t_2)=\fr{\delta^2}{4(R_1^2-R_2 ^2)^2}[R_1^4(t_1^4-t_1^6)+R_2^4(t_2^4-t_2^6)]$ have been used. 
Finally we have
 \begin{equation}
 Res_{s \rightarrow -\half} h(s)=\fr{15\pi}{128}\fr{\delta^2(R_1+R_2)}{(R_1^2-R_2^2)^2}
 \end{equation}
 and therefore
 \begin{equation}
 a_{5/2}= \fr{5\pi^{3/2}}{2} \fr{\delta^2(R_1+R_2)}{(R_1^2-R_2^2)^2}
 \label{coeff_a52}
 \end{equation}
This is the heat kernel coefficient $a_{5/2}$ for the configuration of
the magnetic background field as given by
Eqs.(\ref{potential_A}-\ref{h_r1}).
  
We can calculate the heat kernel coefficient $a_{5/2}$ in a more
general situation when the magnetic field jumps on an {\it arbitrary}
surface $\Sigma$. The coeffcients $a_n$ for $n=1/2,...,2$ can be read
off rather general expressions of the paper \cite{gkv}.  Let $B^\pm$
be values of the magnetic field on two sides of $\Sigma$.  According
the analysis of \cite{gkv}, the coeffcient $a_{5/2}$ must be an
integral over $\Sigma$ of a local invarinat of canonical mass
dimension 4, which is symmetric under the exchange of $B^+$ and $B^-$
and which vanishes if $B^+=B^-$ (i.e. when the singularity
disappears). There is only one such invariant which gives rise to the
following expresssion:
 \begin{equation}
   a_{5/2}= \xi \int\limits_{\Sigma} (\vec{B}^+ -  \vec{B}^-)^2 d\mu(\Sigma),
   \label{general_a52}
 \end{equation}
where the integration goes over the surface $\Sigma$ and
$\vec{B}^{\pm}$ are the values of the magnetic field on both sides of
$\Sigma$ in the given point. The yet undefined constant $\xi $ can be
found using Eq.(\ref{coeff_a52}) which constitutes a special case of
(\ref{general_a52}). Here the surface $\Sigma$ consists of two circles
in the $ (\vec{X},\vec{Y})$-plane so that
 \begin{equation}
 \int\limits_{\Sigma} (\vec{B}^+ -  \vec{B}^-)^2 d\mu(\Sigma)=2\pi(R_1+R_2)\vec{B}^2,   
 \end{equation}
where the jump $(\vec{B}^+ - \vec{B}^-)$ is just the value of
$\vec{B}$ at $r \in [R_1,R_2]$. With this, Eq.(\ref{general_a52})
takes the form
\begin{equation}
   a_{5/2}=2\pi e^2 \xi\vec{B}^2(R_1+R_2).
 \label{agen}
\end{equation} 
On the other side, from Eq.(\ref{fieldb}) we have
\begin{equation}
 \Phi= B\int\limits_{r \in[R_1,R_2]}B d^2 x=\pi B(R_2^2-R_1^2)
\label{phiofb}
\end{equation}
and with Eq.(\ref{delta}) from Eq.(\ref{coeff_a52}) it follows that
 \begin{equation}
 a_{5/2}= \fr{5}{8\sqrt{\pi}} e^2 \Phi^2 \fr{R_1+R_2}{(R_2^2-R_1^2)^2}.
 \label{asp}
 \end{equation}
comparing (\ref{agen}) with (\ref{asp}) we get
\begin{equation}
  \xi = \fr{5}{16\pi^{3/2}}.
\label{ksi}
\end{equation}  

\section{Graphics and numerical results}
\begin{figure}[h]
 \begin{picture}(60,100)(0,15) 
\epsfxsize=7cm\epsffile{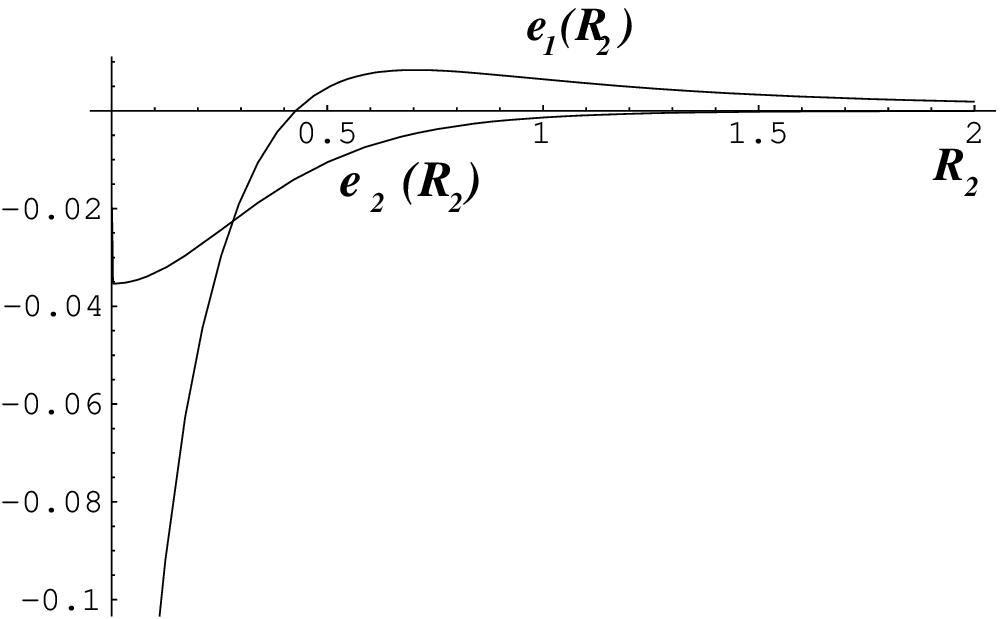}
\epsfxsize=7cm\epsffile{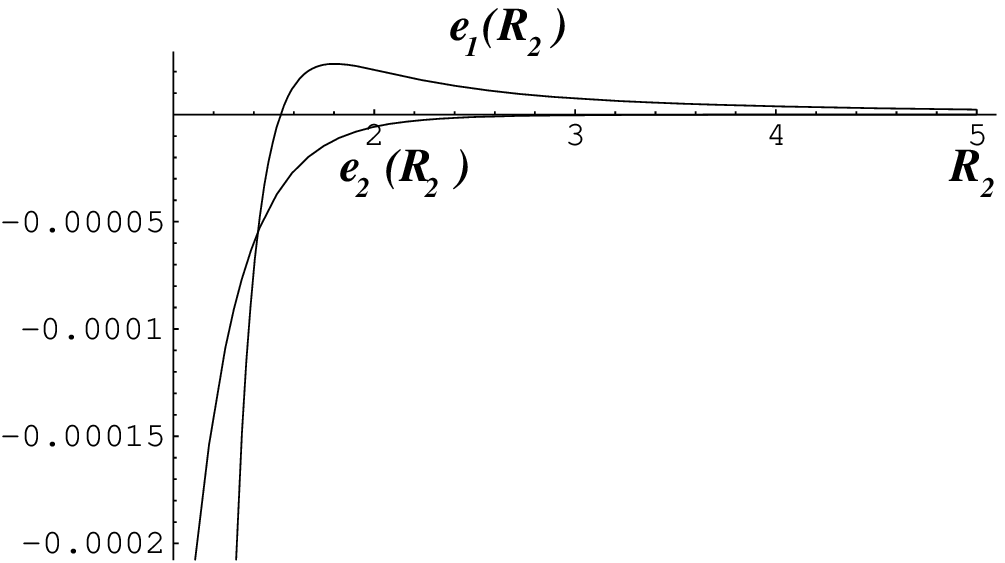} 
 \end{picture} 
\caption{The asymptotical part of energy as function of the outer
radius for $R_1=0.00001$ resp. $R_1=1$ }\label{fig1}
\end{figure}

The asymptotic part of the ground state energy as given by
Eqs. (\ref{Eas}) and (\ref{eas_of_g}) can be represented as sum of a
part proportional to the second power of the flux and one proportional
to the fourth power,
\begin{equation}
E^{as}= -\fr{4}{\pi}[e_1(R_1,R_2)\delta^2+e_2(R_1,R_2)\delta^4]
\label{eas_of_e1_e2}.
\end{equation}
The corresponding coefficients $ e_1(R_1,R_2)$ and $e_2(R_1,R_2)$ can
be calculated numerically without problems. They are shown as
functions of $R_2$ for fixed $R_1$ in Fig.\ref{fig1}  (multiplied by
$R_2^2$). 

\begin{figure}[h] 
\epsfxsize=10cm\epsffile{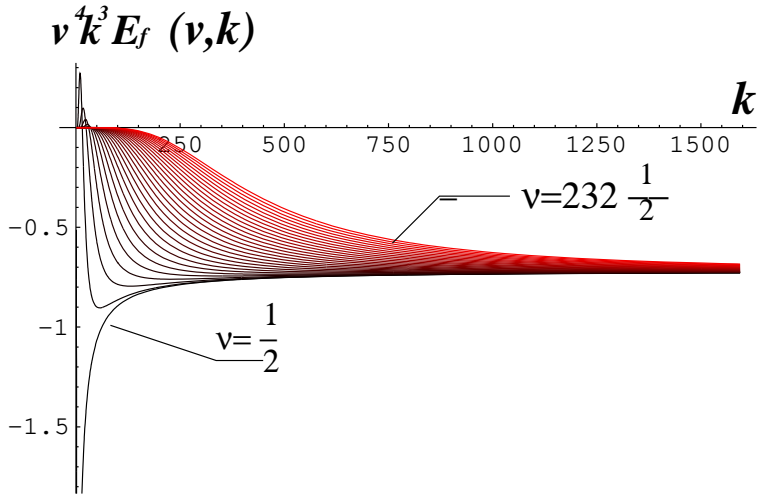}
\caption{The contribution $E_f(\nu,k)$ multiplied by $\nu^4 k^3$ of the
individual radial momenta for several orbital momenta to the finite
part of the ground state energy for $R_1=0.0001$, $R_2=1$, $\delta=3$.}
 \label{fig2}
\end{figure}

\begin{figure}[h]   
\epsfxsize=10cm\epsffile{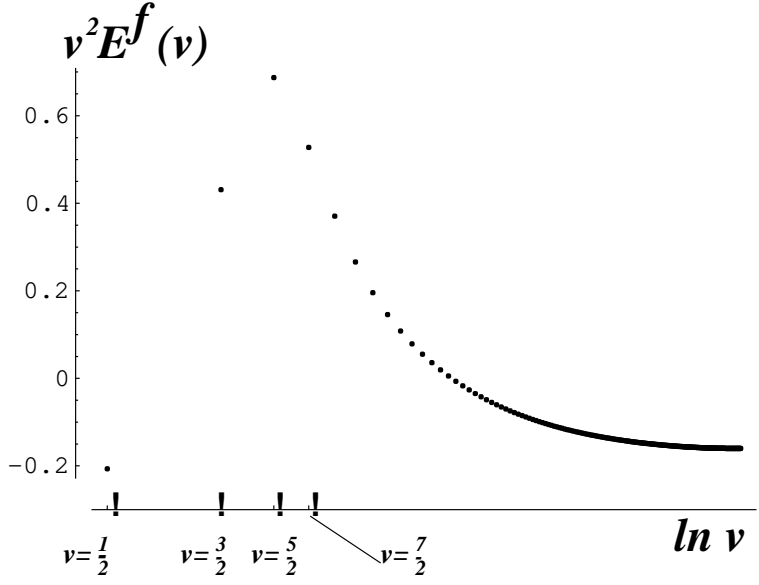}
\caption{The contribution $E_f(\nu)$ of the individual orbital momenta
to the finite part of the ground state energy multiplied by $\nu^2$
for $R_1=0.0001$,$R_2=1$ and $\delta=3$.} \label{fig3}
\end{figure}

The finite part of the ground state energy, $E_f$, is used in the form
as given by Eq. (\ref{ef1}). Let $E_f(\nu,k)$ denote the function to
be integrated and summed over in that expression. In Fig.\ref{fig2} it
is shown as function of $k$ for several values of the orbital momentum
$\nu$. In order the make the behavior better visible it is multiplied
there by $\nu^4 k^3$. These functions are smooth for all values of
$k$, starting from some finite values at $k=0$. For large $k$, the
functions $\nu^4 k^2 E_f(\nu,k)$ shown here tend to a constant thus
the integral over $k$ is convergent.  All integrals have been
truncated at $k=1500$. The error caused by this is quite small and
does not change the results shown in the Table 1. These integrals we
denote by $E_f(\nu)$. They are shown as function of $\nu$ in
Fig.\ref{fig3} in a logarithmic scale. Again, we multiplied by a power
of argument, here by $\nu^2$, in order to make the behavior for large
$\nu$ visible. It is seen that $\nu^2 E_f(\nu)$ tends to a constant so
that the sum over $\nu$ is convergent. The sum is taken up to
$\nu=232.5$ and again the remainder is small.
The calculations have been performed for several values of the
parameters. The results are displayed in Table.1. The computations are
performed with an adapted arithmetical precision. In intermediate
steps compensations between sometimes very large quantities
appeared. The precision was adapted accordingly. For example, for
$R_1=0.99,\nu=250.5,k=1600$ as much as 1404 decimal positions have been
necessary to get at least 16 digits precision of the integrand $E_f(\nu,k)$.
This was a factor causing large computation time.

Table 1. The numerical evaluations for several values of $R_1$ and
$\delta$\\ \\ $R_1$=0.0001\\
\begin{tabular}{c||c|c|c|c|c}\hline
$\delta$ & $E^{f}$ & $E^{as}$ & $E^{ren}$ & $E^{class}$ & $E^{tot}$\\
\hline\hline
0.5& -0.0130152& 0.00152886& -0.0114863& 3.14159& 3.13011\\
1.& -0.052403& 0.00509591& -0.0473071& 12.5664& 12.5191\\
3.& -0.450093& -0.052012& -0.502105& 113.097& 112.595\\
6.& -0.91953& -1.52936& -2.44889& 452.389& 449.94\\
10.& 4.85954& -12.9483& -8.08871& 1256.64& 1248.55\\
15.& 46.3333& -67.3661& -21.0328& 2827.43& 2806.4\\
21.& 215.273& -261.526& -46.2534& 5541.77& 5495.52\\
30.& 987.62& -1095.29& -107.666& 11309.7& 11202.1\\
\hline \hline
\end{tabular}\\
\\
$R_1$=0.9\\
\begin{tabular}{c||c|c|c|c|c}\hline
$\delta$ & $E^{f}$ & $E^{as}$ & $E^{ren}$ & $E^{class}$ & $E^{tot}$\\
\hline\hline
0.5 &-0.278679 &-0.00105651 &-0.279736 & 16.5347 & 16.255  \\ 
1. &-1.11431 &-0.00451479 &-1.11882 &66.1388 & 65.02\\
3. &-10.0633 &-0.0683525 &-10.1317 &595.249 &  585.117\\
6. &-40.4372 & -0.647622&-41.0848 &2381. &  2339.92\\
10. &-112.17 &-4.2629 &-116.433 &6613.88 &  6497.45\\
20. &-423.279 &-63.2506 &-486.53 &26455.5 & 25969.\\
40. &-1079.05 &-992.187 &-2071.24 &105822. & 103751.\\
\hline \hline
\end{tabular}\\
\\
$R_1$=0.95\\
\begin{tabular}{c||c|c|c|c|c}\hline
$\delta$ & $E^{f}$ & $E^{as}$ & $E^{ren}$ & $E^{class}$ & $E^{tot}$\\
\hline\hline
0.5& -0.72166& -0.00212769& -0.723788& 32.2215& 31.4977\\
1.& -2.86633& -0.00877711& -2.87511& 128.886& 126.011\\
3.& -25.8293& -0.104564& -25.9339& 1159.97& 1134.04\\
10.& -287.834& -4.39364& -292.228& 12888.6& 12596.4\\
\hline \hline
\end{tabular}\\
\\
$R_1$=0.99\\
\begin{tabular}{c||c|c|c|c|c}\hline
$\delta$ & $E^{f}$ & $E^{as}$ & $E^{ren}$ & $E^{class}$ & $E^{tot}$\\
\hline\hline
0.5& -5.59211& -0.0106385& -5.60275& 157.869& 152.266\\
1.& -23.2975& -0.0428038& -23.3403& 631.476& 608.136\\
3.& -202.057& -0.409235& -202.466& 5683.28& 5480.82\\
\hline \hline
\end{tabular}\\
\\
$R_1$=0.997\\
\begin{tabular}{c||c|c|c|c|c}\hline
$\delta$ & $E^{f}$ & $E^{as}$ & $E^{ren}$ & $E^{class}$ & $E^{tot}$\\
\hline\hline
1.& -91.996& -0.142036& -92.138& 2097.54& 2005.4\\
\hline \hline
\end{tabular}\\
\\
$R_1$=0.999\\
\begin{tabular}{c||c|c|c|c|c}\hline
$\delta$ & $E^{f}$ & $E^{as}$ & $E^{ren}$ & $E^{class}$ & $E^{tot}$\\
\hline\hline
1.& -311.182& -0.425555& -311.608& 6286.33& 5974.72\\
\hline \hline
\end{tabular}

\section{ Conclusions and Discussions }
\begin{figure}[h] 
\epsfxsize=10cm\epsffile{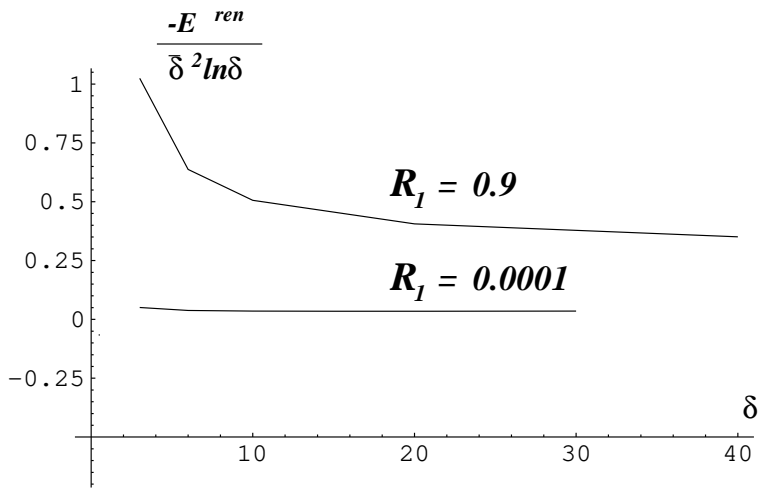}
\caption{The ground state energy divided  by $\delta^2\ln\delta$  as function of $\delta$.}\label{fig4}
\end{figure}

In the preceding sections the ground state energy for a spinor in the
background of a rectangular shaped flux tube had been numerically
calculated. The corresponding Jost function had been written down
explicitly, also its asymptotic part. The numerical calculation
 required work with high arithmetic precision. The results
are displayed mainly in Table 1. For small inner radius of the flux
the results are close to them of \cite{borkir2} where the same problem
for a flux tube with homogeneous magnetic field inside, which
corresponds to $R_1=0$ here, was considered. Especially, it is seen
that for large flux $\delta$ there is a compensation of the
$\delta^4$-contribution between the finite and the asymptotic parts
of the ground state energy leaving a behaviour proportional to
$\delta^2\ln\delta$ as shown in Fig.\ref{fig4}. Here the asymptotic
part gave an essential contribution. The ground state energy remains
negative, but numerically small. Only for very large flux it could
overturn the corresponding classical energy, but these values of the
flux are clearly unphysical.

\begin{figure}[h] 
\epsfxsize=12cm\epsffile{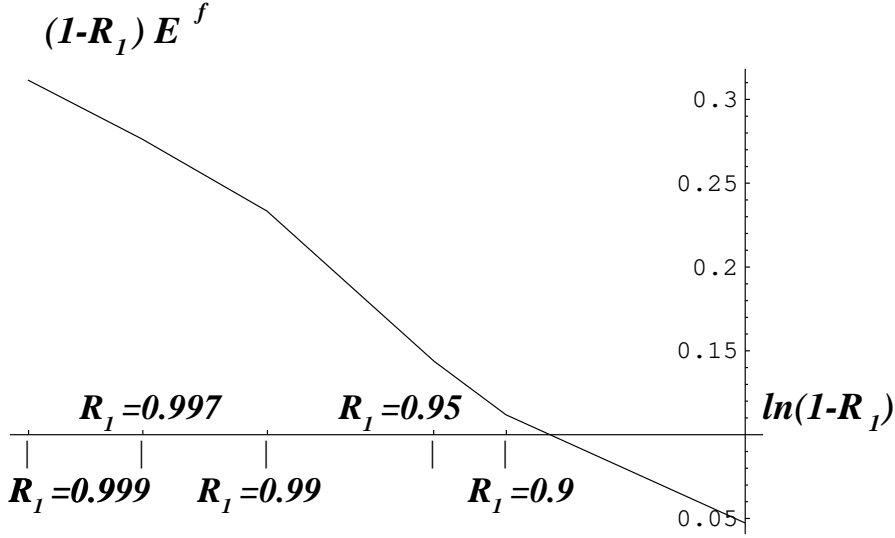}\ \ 
\caption{The vacuum energy multiplied by $(1-R_1)$ in a logarithmic
scale for $\delta=1$.}\label{fig5}
\end{figure}

For values of the inner radius close to the outer one, $R_1\to1$
(where we have put $R_2=1$) the picture changes. Here the vacuum
energy grows faster than the classical one. Generally, both diverge
proportional to $(1-R_1)^{-1}$, the classical energy is equal to
$E_{\rm class}=\delta^2 2\pi (1-R_1)^{-1}$. The vacuum energy,
multiplied by $(1-R_1)$, is shown in Fig.\ref{fig5} in a logarithmic
scale. It is negative and growing a bit faster than the classical one
which would be a constant in this plot. Here the asymptotic part of
the ground state energy becomes increasingly unimportant (see Table 1). The question
whether the vacuum energy for sufficiently small $(1-R_1)$ may become
larger than the classical one cannot be answered by the numerical
results obtained. The problem is that the computations become very
time consuming because of the increasing precision which is
required. Also, one has to take higher $k$ and $\nu$ into account.
The weakening of the growth for $R_1=0.997$ and $R_1=0.999$ seen in
Fig.\ref{fig5} may be caused just by this reason. Here one has to note
that the integrand is for large $k$ and $\nu$ always negative (see
Fig.\ref{fig2}) so that dropping some part (as we did within the
numerical procedure) diminishes the result. So, as a result, we cannot
exclude from the given calculation that the vacuum energy grows for a
strong background field faster than the classical energy.

Further work is necessary to better understand these questions. An
improvement of the numerical procedure is certainly desirable. It
could go along two lines. First, in the calculation of the Jost
function the compensation of large exponentials should be avoided by
taking them into account analytically. Second, in the compensation
between the logarithm of the Jost function and its asymptotic
expansion in the integrand of $E_f$ in Eq.(\ref{ef},\ref{ef1}) higher
orders of the asymptotic expansion could be used. However, for this
reason one would have to continue the procedure invented in
\cite{borkir2} for this expansion using the Lippman-Schwinger equation
or to find some equivalent procedure.

\section*{Acknowledgments}
I thank M. Bordag for advice.\\
I thank D.V. Vassilevich and K. Kirsten for valuable discussions and
helpful remarks.\\
I thank the Graduate college {\it Quantenfeldtheorie} at the University
of Leipzig for support and friendly environment.

\section{Appendix}

 The sum over $v$ has been transformed to integrals
 using the Abel-Plana formula as follows:
\begin{equation}
 \sum\limits_{l=0}^{\infty}(l+\half)=\int\limits_{0}^{\infty}d\nu
f(\nu)+\int\limits_{0}^{\infty}\fr{d\nu}{1+e^{2\pi\nu}}
 \fr{f(i\nu)-f(i\nu)}{i}
 \label{abelplan}
\end{equation}
The integrations over $\nu$ and $k$ can be done using identities:
\begin{equation}
\int\limits_{0}^{\infty} d\nu \int\limits_{m}^{\infty}
dk(k^2-m^2)^{1-s}\fr{\pd}{\pd k}\fr{t^j}{\nu^n}= -\fr{m^{2-2s}}{2}
\fr{\Gamma(2-s)\Gamma(\fr{1+j-n}{2})\Gamma(s+\fr{n-3}{2})}{(rm)^{n-1}\Gamma(\fr{j}{2})}
\label{ident1}
\end{equation}

\begin{equation}
 \int\limits_{m}^{\infty}
dk(k^2-m^2)^{1-s}\fr{\pd}{\pd k}t^j= -m^{2-2s}
\fr{\Gamma(2-s)\Gamma(s+\fr{j}{2}-1)}{\Gamma(\fr{j}{2})}\fr{(\fr{\nu}{rm})^j
 }{(1+(\fr{\nu}{rm})^2)^{s+\fr{j}{2}-1}}
\label{ident2}
\end{equation}
The expansion in powers of $\nu$ for logarithm of asymptotic Jost function
can be obtained in the form (see\cite{borkir1})\\
\begin{equation}
ln \boldmath f_{as}\unboldmath(ik)=\sum\limits_{n=1}^3\sum\limits_{j=1}^9
\int\limits_{0}^{\infty} \fr{dr}{r}X_{nj}\fr{t^j}{\nu^n}\\
\label{lnfas_x}
\end{equation}
\\
where\\
$X_{1,1}= -X_{1,3} =X_{2,6} =\half (a\delta)^2\\$\\
$X_{2,2}= \fr{1}{4}\delta^2(a^2-raa')\\$\\
$X_{2,4}= \fr{1}{4}\delta^2 (-3a^2+raa')\\$\\
$X_{3,3}= \fr{1}{4}\delta^2(a^2-raa'+\half r^2 aa''-\half \delta^2 a^4 )\\$\\
$X_{3,5}= \fr{1}{8}\delta^2(-\fr{39}{2}a^2+7raa'-r^2aa''+6\delta^2 a^4 )\\$\\
$X_{3,7}= \fr{1}{8}\delta^2(35 a^2 -5raa'-5\delta^2 a^4 )\\$\\
$X_{3,9}= \fr{-35}{16}\delta^2 a^2\\$\\

 For the representation of $E_{as}$ as \ref{eas_of_g}
\begin{equation}
E^{as}= \fr{-4}{\pi}\int\limits_{0}^{\infty}\fr{dr}{r^3}[\delta^2 a(r)^2
g_1(rm)-\delta^2 r^2 a'(r)^2 g_2(rm)+\delta^4 a(r)^4 g_3(rm)]\\
\label{eas_of_g_f}
\end{equation}
here $f_i$ are
\begin{eqnarray}
f_1(x)&=& \half f_{1, 1}(x)-\half f_{1, 3}(x)+\fr{1}{4}f_{3,
  3}-\fr{39}{16}f_{3, 5}(x)+\fr{35}{8}f_{3, 7}(x)-\fr{35}{16} f_{3,
  9}(x)\nn\\&-&
\half x \pd_x( -\fr{1}{4} f_{3, 3}(x)+\fr{7}{8}f_{3, 5}(x)-\fr{5}{8}f_{3, 7}(x))\nn\\
&+&\half x \pd_x^2(\fr{x}{8}f_{3, 3}(x)- \fr{x}{8}f_{3, 5}(x) ), 
\nn\\
 f_2(x)&=& \fr{1}{8}(f_{3, 3}(x) - f_{3, 5}(x)), \nn\\
 f_3(x)&=& -\fr{1}{8}(f_{3, 3}(x) - 6 f_{3, 5}(x)+5 f_{3, 7}(x) ). 
\end{eqnarray}
with $f_{i. j}$ are

\begin{eqnarray}
f_{1, 1}(x)&=& - \fr{1}{1+e^{2\pi x}}\nn\\
f_{1, 3}(x)&=& - (\fr{}{1+e^{2\pi x}})' \nn\\
f_{3, 3}(x)&=& (\fr{1}{x} \fr{1}{1+e^{2\pi x}} )'\nn\\
f_{3, 5}(x)&=& \fr{1}{3}( \fr{1}{x}( \fr{x}{1+e^{2\pi x}})')'\nn\\
f_{3, 7}(x)&=& \fr{1}{15}( \fr{1}{x}( \fr{1}{x}( \fr{x^3}{1+e^{2\pi x}} )')')'\nn\\
f_{3, 9}(x)&=& \fr{1}{105}( \fr{1}{x}( \fr{1}{x}( \fr{1}{x}(
\fr{x^5}{1+e^{2\pi x}})')')')'
\label{f_i_of_v}
\end{eqnarray}
The asymtotic of the logarithmic Jost function can be obtained in the form:
$\ln f_{\nu}(ik)= \sum\limits_{n}h_n(t_1,t_2) $,(see \ref{unif_as_f})
with:
\\
\begin{eqnarray}
h_1&=&\frac{4}{3}\lambda^2[
{\Mvariable{R1}}^4 \Mvariable{t1}( 2 + \Mvariable{t1}) {( 1 + \Mvariable{t2}) }^2
-3{\Mvariable{R1}}^2{\Mvariable{R2}}^2 {(1 +\Mvariable{t1}) }^2 \Mvariable{t2} ( 1 + \Mvariable{t2})+
{\Mvariable{R2}}^4 { ( 1 + \Mvariable{t1}) }^2\Mvariable{t2}( 1 +
  2\Mvariable{t2})
  ]\nn\\
\nn\\
h_2&=&0\nn\\
\nn\\
h_3&=&-\frac{1}{6}{\lambda}^2 [ 2{\Mvariable{R1}}^4{\Mvariable{t1}}^3 {( 1 + \Mvariable{t2}) }^2 +{\Mvariable{R1}}^2 {\Mvariable{R2}}^2
{( 1 + \Mvariable{t1}) }^2 {\Mvariable{t2}}^3{( 1 + \Mvariable{t2}) }^2(3 {\Mvariable{t2}}^2 - 4)-\nn\\&&
{\Mvariable{R2}}^4 {( 1 + \Mvariable{t1}) }^2{\Mvariable{t2}}^3
(3{\Mvariable{t2}}^4 + 6 {\Mvariable{t2}}^3 -{\Mvariable{t2}}^2 -
8\Mvariable{t2} - 2 ) ]+\nn\\
&& \frac{2}{15} {\lambda}^4 [ 4 {\Mvariable{R1}}^8
{\Mvariable{t1}}^3( 4 + \Mvariable{t1}){( 1 + \Mvariable{t2}) }^4-
5 {\Mvariable{R1}}^6 {\Mvariable{R2}}^2{( 1 + \Mvariable{t1}) }^4
{\Mvariable{t2}}^3{( 1 + \Mvariable{t2}) }^4 +\nn\\
&&15 {\Mvariable{R1}}^4 {\Mvariable{R2}}^4 {( 1 + \Mvariable{t1}) }^4 {\Mvariable{t2}}^3 {( 1 +
  \Mvariable{t2} ) }^2 ( {\Mvariable{t2}}^2 + 2 \Mvariable{t2} - 1)-\nn\\&&
5 {\Mvariable{R1}}^2 {\Mvariable{R2}}^6 { ( 1 + \Mvariable{t1}) }^4
{\Mvariable{t2}}^3( 3 {\Mvariable{t2}}^4 +12{\Mvariable{t2}}^3 + 6
{\Mvariable{t2}}^2 - 4 \Mvariable{t2} - 1 )+\nn\\&&
{\Mvariable{R2}}^8 {( 1 + \Mvariable{t1} ) }^4{\Mvariable{t2}}^3 (5
{\Mvariable{t2}}^4 +20 {\Mvariable{t2}}^3 - 4\Mvariable{t2} -1 ) ]\nn\\
\nn\\
h_4&=&\frac{\delta^2 [ {\Mvariable{R1}}^4 {\Mvariable{t1}}^4 ( 1 - {\Mvariable{t1}}^2)  + 
         {\Mvariable{R2}}^4 {\Mvariable{t2}}^4 ( 1 - {\Mvariable{t2}}^2 ) ]}
          {4 {( \Mvariable{R1}^2 - \Mvariable{R2}^2) }^2 }\nn\\
\nn\\
\lambda&=&  \frac{\delta}{( \Mvariable{R1}^2 - \Mvariable{R2}^2 )
     ( 1 + \Mvariable{t1}) ( 1 + \Mvariable{t2})}
 \label{unif_as_h}
\end{eqnarray}   
\\

\end{document}